\documentclass[12pt,a4paper]{article}
\usepackage{graphicx,rotating,epsfig,amsmath,amssymb,color,multirow,pstricks,pst-plot,comment}
\newcommand{\cinst}[2]{$^{\mathrm{#1)}}$~#2\par}
\newcommand{\crefi}[1]{$^{\mathrm{#1)}}$}
\newcommand{\HRule}{\rule{0.4\linewidth}{0.3mm}}

\begin{document}

\begingroup
\thispagestyle{empty} \baselineskip=14pt
\parskip 0pt plus 5pt

\begin{center}
{\large EUROPEAN ORGANIZATION FOR NUCLEAR RESEARCH}
\end{center}

\bigskip
\begin{flushright}
~\\
March 29, 2013
\end{flushright}

\bigskip
\begin{center}
{\Large\bf \boldmath
Minimal physical constraints on the\\[3mm] 
angular distributions of two-body boson decays
}

\bigskip\bigskip

Pietro Faccioli\crefi{1,2},
Carlos Louren\c{c}o\crefi{3},
Jo\~ao Seixas\crefi{1,2}
and Hermine K.\ W\"ohri\crefi{3}

\bigskip\bigskip\bigskip
\textbf{Abstract}

\end{center}

\begingroup
\leftskip=0.4cm \rightskip=0.4cm \parindent=0.pt

The angular distribution of the two-body decay of a boson of unknown
properties is strongly constrained by angular momentum conservation and rotation
invariance, as well as by the nature of the detected decay particles and of the
colliding ones. Knowing the border between the ``physical''
and ``unphysical'' parameter domains defined by these ``minimal constraints"
(excluding specific hypotheses on what is still subject of measurement) is
a useful ingredient in the experimental determinations of angular
distributions and can provide model-independent criteria for spin characterizations.
In particular, analysing the
angular decay distribution with the general parametrization for the $J = 2$ case
can provide a model-independent discrimination between
the $J=0$ and $J=2$ hypotheses for a particle produced
by two real gluons and decaying into two real photons.
\bigskip
\endgroup
\vspace{1cm}
\begin{center}
\end{center}
\vfill
\begin{flushleft}
\HRule\\

\cinst{1} {Laborat\'orio de Instrumenta\c{c}\~ao e F\'{\i}sica Experimental de Part\'{\i}culas (LIP),\\ ~~~
Lisbon, Portugal}
\cinst{2} {Physics Department, Instituto Superior T\'ecnico (IST), 
Lisbon, Portugal}
\cinst{3} {European Organization for Nuclear Research (CERN), 
Geneva 23, Switzerland}
\end{flushleft}
\endgroup

\newpage

\sloppy

\section{Introduction} \label{sec:intro}

Measurements of the angular distributions of particle decays give unique
insights into the underlying fundamental interactions and play a central role
in the determination of coupling properties, in the verification of production
models and even in the discovery and identification of new particles.

However, some of the most basic properties of the decay distributions are
ignored in many experimental analyses of Standard-Model (SM) couplings of vector bosons,
Drell--Yan and quarkonium production. Only recently some general
characteristics of the angular distribution have been systematically addressed,
highlighting the importance of the choice of polarization
axis~\cite{bib:framedep1,bib:framedep2,bib:EPJC}, 
revealing the existence of general frame-independent
relations~\cite{bib:lamdatilde,bib:PVpaper} and precisely defining the shape of
the allowed parameter domain~\cite{bib:dilepton_constraints}.

In this paper we determine the physically-allowed parameter
domains of the two-body decays of bosons, generalizing a previous study
limited to the parity-conserving dilepton decays of $J=1$ particles~\cite{bib:dilepton_constraints}.
Considering a broad range of physical processes involving two-body decays, we
derive model-independent constraints that the parameters of the decay
distributions must obey. These ``minimal physical constraints'' (MPCs) are
determined only by angular momentum conservation, by rotation invariance and by
the identities of the initial- and final-state particles, minimizing hypotheses
on the nature and properties of the decaying particle.

A very interesting by-product of this study is the observation that $J = 0$ and $J = 2$ 
bosons produced by two real gluons and decaying into two real photons have 
well-separated physically-allowed parameter spaces.
Therefore, these two spin options can be discriminated with a 
model-independent procedure, describing the measured angular decay distribution 
using the general parametrization for the $J=2$ case and studying the extracted 
anisotropy parameters (only two in the case of the polar projection) as
contours in the parameter space.

The paper is organized as follows.
After discussing the formalism used for the description of the two-body decay
angular distribution (Sec.~\ref{sec:genericJ}) and, in particular, its polar
projection (Sec.~\ref{sec:polar}), we address in detail the case of a $J=1$
particle (Sec.~\ref{sec:J1}), 
considering the most general two-body decay as well as, specifically,
the decay into a fermion-antifermion pair, including possible parity-violating
effects. In Sec.~\ref{sec:J2} we discuss the $J=2$ case, focusing on the polar
projection of the decay distribution and giving examples of how the MPCs affect
the angular parameters for different production and decay modes. Finally, in
Sec.~\ref{sec:J_eq_0_verification} we illustrate the utility of the MPCs in
analyses aimed at characterizing the spin of new particles, such as
the heavy di-photon resonance recently discovered at
CERN~\cite{bib:ATLAS,bib:CMS}.

\section{The two-body decay distribution of a boson} \label{sec:genericJ}

\begin{figure}[t]
\centering
\includegraphics[width=0.5\linewidth]{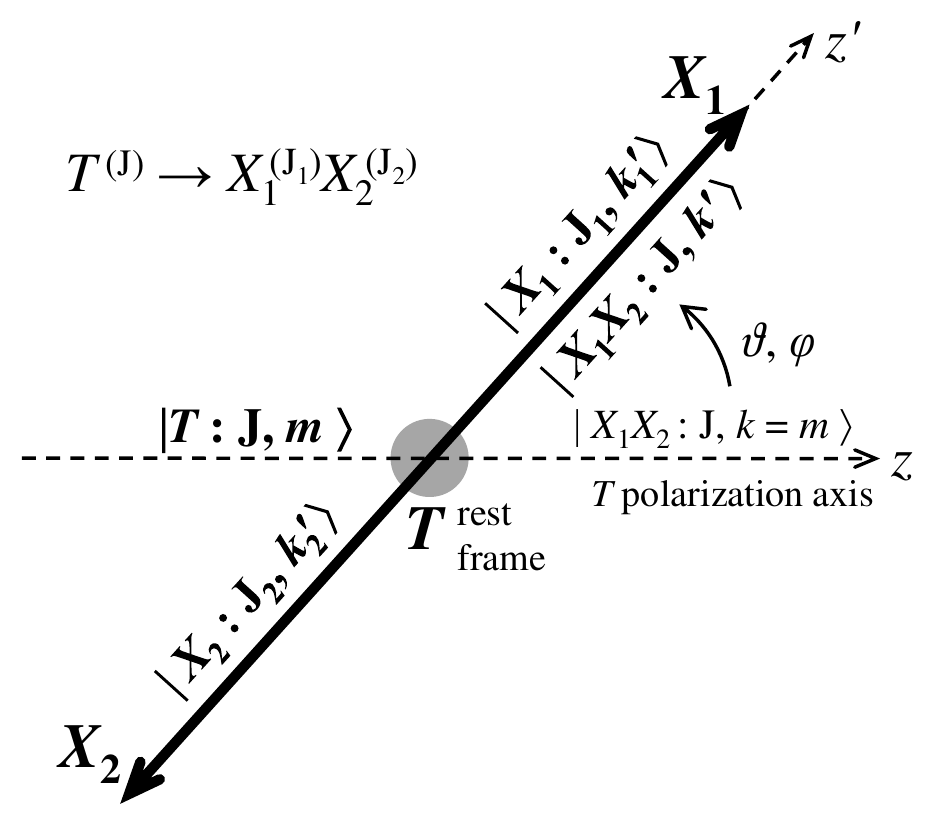}
\caption{\label{fig:defs} Sketch of the $T \rightarrow X_1 X_2$ decay,
specifying notations for axes, angles and angular momentum states.}
\end{figure}

Using the notations defined in Fig.~\ref{fig:defs}, the decay
of a particle $T$ of total angular momentum quantum number $J$ into
two observed particles $X_1$ and $X_2$ of angular momenta $J_1$ and
$J_2$ has the following angular distribution:
\begin{align}
& 
W(\cos\!\vartheta, \varphi) = 
\label{eq:most_general_ang_distr}
\\
&
\sum_{\begin{smallmatrix} |k^\prime| \le \min(J,J_1+J_2) \\ |m| \le J, \; |n| \le J \end{smallmatrix}}
 \rho_{m, n}^{k^\prime} \;
\mathcal{D}_{m, k^\prime}^{J}(\cos\!\vartheta, \varphi) \mathcal{D}_{n,
k^\prime}^{J*}(\cos\!\vartheta, \varphi)  \, ,
\nonumber 
\end{align}
with
\begin{equation}
\rho_{m, n}^{k^\prime} \; = \sum_{\begin{smallmatrix} |k_1^\prime| \le J_1, \;
|k_2^\prime| \le J_2 \\ k_1^\prime + k_2^\prime = k^\prime \end{smallmatrix}}
\left\langle \mathcal{A}_{m, k_1^\prime, k_2^\prime}^* \mathcal{A}_{n,
k_1^\prime, k_2^\prime} \right\rangle \, . \label{eq:rho}
\end{equation}

In these equations,
$k_1^\prime$ and $k_2^\prime$ are the angular momentum projections of $X_1$ and
$X_2$ along their common direction in the $T$ rest frame ($z^\prime$ axis), and
$m$ is the $T$ angular momentum projection along the chosen
$T$ quantization axis $z$ (for example the $T$ momentum in the laboratory, or
the direction of one or the other colliding beam, or the average of the two). The
angular functions $\mathcal{D}_{m, k^\prime}^{J}$ are the elements of the
Wigner matrix, which applies the axis rotation $xyz \to x^\prime y^\prime
z^\prime$ to the $X_1 X_2$ angular momentum state, re-expressing the pure
$J_{z^\prime}$ eigenstate $|X_1 X_2: J, k^\prime \rangle$ into a combination of
$J_z$ eigenstates $|X_1 X_2: J, k \rangle$. The rotation angles are the
(polar and azimuthal) decay angles $\vartheta$ and $\varphi$, defined as the
angles formed by $X_1$ (or by $X_2$) with respect to the axis $z$ and a
suitably defined $xz$ plane (for example the ``production plane'', containing the momenta of the
colliding beams and of the decaying particle).

The amplitudes $\mathcal{A}_{m, k_1^\prime, k_2^\prime}$, depending on the
elementary couplings in the production and decay process considered, determine
the allowed combinations of initial- and final-state ``helicities'' (here we
define the helicity of a particle as its total angular momentum projection along a
specified axis). For example, in the decays of SM vector gauge
bosons into fermions, all amplitudes with opposite fermion spin projections,
$\mathcal{A}_{m, \pm 1/2, \mp 1/2}$, vanish in the limit of massless fermions
because of helicity conservation. In Eq.~\ref{eq:rho}, the amplitude products
$\mathcal{A}_{m, k_1^\prime, k_2^\prime}^* \mathcal{A}_{n, k_1^\prime,
k_2^\prime}$ appear in a weighted average over all subprocesses contributing to
the production of $T$, with weights proportional to their relative yields.
The complex coefficients $\rho_{m, n}^{k^\prime}$ form, for each
$k^\prime$, a hermitian matrix ($\rho_{m, n}^{k^\prime} = \rho_{n, m}^{k^\prime*}$) 
with trace unity, often referred to as ``spin density matrix''.

It is worth reminding that the relation $|J_1 - J_2| \le J \le J_1 + J_2$ is
not necessarily satisfied when the decay products $X_1$ and $X_2$ have a
relative orbital angular momentum (for example, a $J=2$ particle can decay into
two fermions with $J_1=J_2=1/2$ and relative orbital angular momentum $L \ge
1$). However, $L$ does not play an explicit role in
the expression of the decay distribution,  Eq.~\ref{eq:most_general_ang_distr},
because its projection along the quantization axis of the decay products
($z^\prime$) is always zero.

\section{The polar projection of the decay distribution} \label{sec:polar}

The polar projection of the general decay angular distribution,
Eq.~\ref{eq:most_general_ang_distr}, is obtained by integrating over $\varphi$.
Given the general identity
\begin{equation}
\int \mathrm{Re} \left[ \mathcal{C} \; \mathcal{D}_{m,
k^\prime}^{J*}(\cos\!\vartheta, \varphi) \mathcal{D}_{n,
k^\prime}^{J}(\cos\!\vartheta, \varphi) \right] d\varphi = 0 \; ,
\label{eq:Dmatrix_identity}
\end{equation}
for $n \ne m$,
where $\mathcal{C}$ is any combination of complex amplitudes not depending on
$\varphi$,
the integrated expression is
\begin{equation}
w(\cos\!\vartheta) \;
 = \; \sum_{\begin{smallmatrix} |k^\prime| \le \min(J,J_1+J_2) \\ |m| \le J \end{smallmatrix}}
\sigma_{m,k^\prime} \left[ d^J_{m,k^\prime}(\cos\!\vartheta) \right]^2 \, ,
\label{eq:ang_distr_polar_projection_general}
\end{equation}
where $d^J_{m,k^\prime}$ are the reduced Wigner matrix elements,
functions of $\cos\!\vartheta$, and
$\sigma_{m,k^\prime} \equiv \; \rho_{m, m}^{k^\prime}$.
%
%
This expression only depends on the squared moduli of the helicity
amplitudes (``diagonal'' $\rho_{m, m}^{k^\prime}$ terms). Therefore, any
information about the interference between different angular momentum
eigenstates composing the initial state is lost in the polar projection of the
distribution.

The decaying particle can be a coherent or an incoherent superposition of
eigenstates (as exemplified in Sec.~4 of Ref.~\cite{bib:EPJC}). 
These two physically different cases lead to different azimuthal
anisotropies, properly reflected in Eq.~\ref{eq:most_general_ang_distr}, but
not to different polar anisotropies, thereby being indistinguishable in the
polar projection. While the maximum number of observable parameters of the full
distribution is 8 for $J=1$ (5 if the particle is observed inclusively, without
referring the polarization axes to possible accompanying particles in the
event) and 24 for $J=2$ (14 for inclusive observation), the corresponding polar
projections have only 2 and 4 measurable parameters, respectively.
The azimuthal dependence of the distribution obviously vanishes when the particle is
produced in $2 \to 1$ processes and the polarization axis $z$ is chosen along
the direction of the colliding particles.

For a generic $J$, the polar angle
projection can be expressed in terms of $2J$ independent observable
coefficients, $\lambda_i$, as
\begin{equation}
w( \cos\!\vartheta \; | \; \vec{\lambda} ) \,
  = \, \frac{1}{2} \, \frac{1 + \sum_{i=1}^{2J} \lambda_i \, (\cos\!\vartheta)^i }
  {1 + \sum_{j=1}^{J} \frac{\lambda_{2j}}{2j+1} } \, .
 \label{eq:polar_ang_distr_anyJ}
\end{equation}

\section{\boldmath The $J=1$ specific case} \label{sec:J1}

The most general form of the two-body decay angular distribution of a $J=1$
particle is
\begin{align}
  W(\cos\!\vartheta, &\varphi) \, = \,
  \frac{3/(4\pi)}{(3 + \lambda_{\vartheta})} \,
  (1 + \lambda_{\vartheta} \cos^2\!\vartheta\  + \nonumber \\
  & + \lambda_{\varphi} \sin^2\!\vartheta \cos\!2\varphi
  + \lambda_{\vartheta \varphi} \sin\!2\vartheta \cos\!\varphi 
  \label{eq:ang_distr_J1}  \\
  & + \lambda^{\bot}_{\varphi} \sin^2\!\vartheta \sin\!2\varphi +
  \lambda^{\bot}_{\vartheta \varphi} \sin\!2\vartheta \sin\!\varphi \nonumber\\
  & + 2A_{\vartheta} \cos\!\vartheta
    + 2A_{\varphi} \sin\!\vartheta \cos\!\varphi
    + 2A^{\bot}_{\varphi} \sin\!\vartheta \sin\!\varphi
  ) \, ,  \nonumber
\end{align}
where
\begin{align}
  \lambda_{\vartheta} =
  & \, 1/ D \, [ 4 \rho_{0,0}^{0} 
  + \rho_{+1,+1}^{+1} + \rho_{+1,+1}^{-1} + \rho_{-1,-1}^{+1} + \rho_{-1,-1}^{-1} \nonumber \\
  & \qquad -2 ( \rho_{0,0}^{+1} + \rho_{0,0}^{-1} + \rho_{+1,+1}^{0} + \rho_{-1,-1}^{0} )] \, , \nonumber \\
  \lambda_{\varphi} =
  & \, 2/D \, \mathrm{Re} ( \rho_{+1,-1}^{+1} + \rho_{+1,-1}^{-1} -2 \rho_{+1,-1}^{0} ) \, , \nonumber \\
  \lambda^{\bot}_{\varphi} =
  & \, 2/D \, \mathrm{Im} ( \rho_{+1,-1}^{+1} + \rho_{+1,-1}^{-1} -2 \rho_{+1,-1}^{0} ) \, , \nonumber \\
  \lambda_{\vartheta \varphi} =
  & \, \sqrt{2}/D \, \mathrm{Re} [ \rho_{+1,0}^{+1} + \rho_{+1,0}^{-1} -2 \rho_{+1,0}^{0} \nonumber \\
  & \qquad - ( \rho_{0,-1}^{+1} + \rho_{0,-1}^{-1} -2 \rho_{0,-1}^{0} ) ] \, , 
      \label{eq:lambdas_J1} \\
  \lambda^{\bot}_{\vartheta \varphi} =
  & \, \sqrt{2}/D \, \mathrm{Im} [ \rho_{+1,0}^{+1} + \rho_{+1,0}^{-1} -2 \rho_{+1,0}^{0} \nonumber \\
  & \qquad - ( \rho_{0,-1}^{+1} + \rho_{0,-1}^{-1} -2 \rho_{0,-1}^{0} ) ] \, , 
  \nonumber \\
  A_{\vartheta} =
  & \, 1/D \, ( \rho_{+1,+1}^{+1} - \rho_{+1,+1}^{-1} +
  \rho_{-1,-1}^{-1} - \rho_{-1,-1}^{+1}  ) \, , \nonumber \\
  A_{\varphi} =
  & \, \sqrt{2} / D \,
  \mathrm{Re} ( \rho_{+1,0}^{+1} - \rho_{+1,0}^{-1} +
  \rho_{0,-1}^{+1} - \rho_{0,-1}^{-1}  )  \, , \nonumber \\
  A^{\bot}_{\varphi} =
  & \, \sqrt{2} / D \,
  \mathrm{Im} ( \rho_{+1,0}^{+1} - \rho_{+1,0}^{-1} +
  \rho_{0,-1}^{+1} - \rho_{0,-1}^{-1}  )  \, , \nonumber \\
  &\mathrm{with} \; \; D =
  \,  \rho_{+1,+1}^{+1} + \rho_{+1,+1}^{-1} + \rho_{-1,-1}^{+1} +
  \rho_{-1,-1}^{-1} \nonumber \\
  &  \qquad\qquad +2 ( \rho_{0,0}^{+1} + \rho_{0,0}^{-1} + \rho_{+1,+1}^{0} + \rho_{-1,-1}^{0} ) \, . \nonumber
\end{align}
The parameters $\lambda_\varphi^{\perp}$, $\lambda_{\vartheta
\varphi}^{\perp}$ and $A_\varphi^{\perp}$ are of physical interest
only if the particle is observed in an exclusive production
channel, taking into account the momenta of accompanying
particles in a specific event topology, in which case at least one other
physical plane, besides the one formed by the colliding beams, can
be chosen as a meaningful reference for the definition of the
azimuthal angle.

The physically observable parameter domain is defined by the following MPCs,
obtained using Eqs.~\ref{eq:rho} and~\ref{eq:lambdas_J1}:
\begin{align}
\lambda_{\vartheta \varphi}^{\otimes 2} + 2 (\tilde{A}^2 + A_\vartheta^2)
& \le  2 (1 + \lambda_\vartheta) \, , \nonumber \\
\lambda_{\vartheta \varphi}^{\otimes 2} 
& \le 2 (1 + \lambda_\vartheta) \left[1 - \sqrt{2
A_{\varphi}^{\otimes 2} }\,\right]   , \nonumber \\
\lambda_{\vartheta \varphi}^{\otimes 2}
& \le 2 (1 + \lambda_\vartheta) (1 - |A_\vartheta| )  \, , \nonumber \\
4 A_\vartheta^2 & \le  (1 + \lambda_\vartheta)^2 \, , \nonumber \\
\tilde{A}^2 + A_{\varphi}^{\otimes 2}
& \le  1 \, , \nonumber
\end{align}
\begin{align}
\tilde{A}^2 + A_{\varphi}^{\otimes 2}
 + \lambda_\vartheta^2 & \le  1  \;\;
\mathrm{if}
\;\; \lambda_\vartheta < 0 \, , \nonumber \\
A_{\varphi}^{\otimes 2} 
& \le (1 + \lambda_\vartheta) \left(1 - \sqrt{A_\vartheta^2 + 
\lambda_{\varphi}^{\otimes 2} }\,\right)  , \nonumber \\
\tilde{A}^2 + \lambda_{\varphi}^{\otimes 2} 
& \le  1 \, ,
\nonumber \\
\lambda_{\vartheta \varphi}^{2} + \tilde{A}^2
& \le (1 + \lambda_\varphi)(1 + \lambda_\vartheta)  \, , \nonumber \\
4 A_\varphi^2 & \le  (1 + \lambda_\varphi)^2 \, , 
\label{eq:constraints_J1_general} \\
2 A_{\varphi}^{2} + \lambda_{\varphi}^{\perp 2} & \le 1 \, , \nonumber \\
2 A_{\varphi}^{2} + (1- 2 \lambda_\varphi)^2  & \le  1 \;\; \mathrm{if} \;\; \lambda_\varphi > 1/3 \, , \nonumber \\
2 A_{\varphi}^{2} + (1-2\lambda_{\varphi})^2 + \lambda_{\varphi}^{\perp 2} &
\le 1 \;\; \mathrm{if} \;\; \lambda_\varphi > 1/2 \, , \nonumber \\
2 A_{\varphi}^{2} + \lambda_{\varphi}^{\otimes 2}
& \le 1 \;\; \mathrm{if} \;\; \lambda_\varphi < 0 \, , \nonumber 
%
\end{align}
where 
\begin{equation}
X_i^{\otimes} \equiv \sqrt{X_i^{2} + X_i^{\perp 2}}
\quad \mathrm{and} \quad 
\tilde{A} \equiv 
\sqrt{A_{\vartheta}^{2} + A_{\varphi}^{2} + A_{\varphi}^{\perp 2}} \, .
\nonumber
\end{equation}
\noindent
The last 6 inequalities also apply after the simultaneous substitutions 
$A_{\varphi} \to A_\varphi^{\perp}$,
$\lambda_{\vartheta \varphi} \to \lambda_{\vartheta \varphi}^{\perp}$
and $\lambda_{\varphi} \to -\lambda_{\varphi}$.

In the case of the decay into a fermion-antifermion pair via an intermediate vector
boson (when $k^\prime = 0$, i.e.\ $\rho^0_{m,n} = 0$, is forbidden because of
helicity conservation, valid in the limit of massless fermions or heavy
initial state), the MPCs can be written as
\begin{align}
\begin{split}
G(\lambda_\varphi,\lambda_\vartheta) & \ge  2 (\lambda_{\vartheta
\varphi}^{\perp 2} + A_{\varphi}^{2}) \, , \\
G(-\lambda_\varphi,\lambda_\vartheta) & \ge  2 (\lambda_{\vartheta
\varphi}^{2} + A_{\varphi}^{\perp 2}) \, , \\
G\left(\sqrt{\lambda_\varphi^{\otimes 2}
+A_\vartheta^2},
\lambda_\vartheta\right) & \ge  2 (\lambda_{\vartheta \varphi}^{\otimes 2} 
+ A_{\varphi}^{\otimes 2} ) \, , \\
G\left(-\sqrt{\lambda_\varphi^{\otimes 2}+A_\vartheta^2},
\lambda_\vartheta\right) & \ge  0 \, , \label{eq:constraints_J1_dilepton}
\end{split}
\end{align}
%
%
%
with $G(X,Y)  =  1/2\,[(1+X)^2 - (Y+X)^2]$.

As a result of the inequalities themselves, the left-side terms ($G$ functions)
are bound between $0$ and $1$, and the right sides of the inequalities are
sums of quantities all individually included between $0$ and $1$. These
inequalities become equivalent to those presented in
Ref.~\cite{bib:dilepton_constraints} when only the three parity-conserving
observables $\lambda_{\vartheta}$, $\lambda_{\varphi}$ and $\lambda_{\vartheta
\varphi}$ are considered. We note that a misprint occurred in Eq.~3 of
Ref.~\cite{bib:PVpaper}, where the factors $\sqrt{2} / (2 \mathcal{D})$
starting the expressions of $A_{\varphi}$ and $A^{\bot}_{\varphi}$ should be
replaced by $\sqrt{2} / \mathcal{D}$. This misprint does not affect any of the
subsequent formulas (frame transformation, frame-invariant asymmetries)
involving $A_{\varphi}$ and $A^{\bot}_{\varphi}$. With this correction, the
expressions of Ref.~\cite{bib:PVpaper}, valid in the special case of di-fermion
decays with fermion helicity conservation, where $\rho_{m,n}^{0} = 0$, are a
particular version of the expressions presented in Eq.~\ref{eq:lambdas_J1}.

\begin{figure}[tb]
\centering
\resizebox{0.65\linewidth}{!}{%
\includegraphics{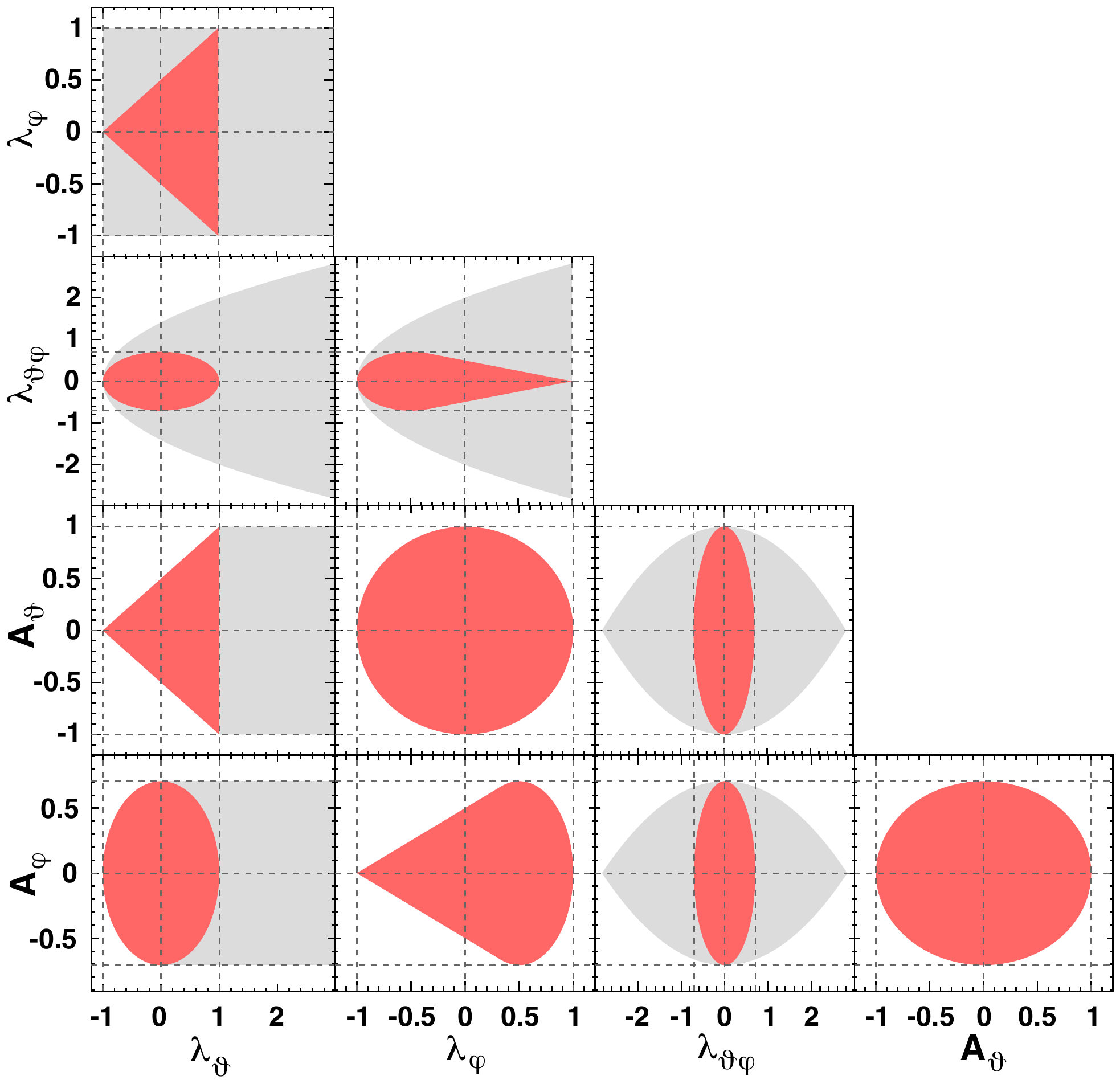}}
\caption{\label{fig:domains_J1} Allowed parameter regions of the
two-body decay distribution of an inclusively observed $J=1$ boson. The
largest areas (gray) represent the most general domain while the inner
areas (red) represent the di-fermion decay case. In
the most general case there are no upper bounds on $\lambda_{\vartheta}$ and
$|\lambda_{\vartheta \varphi}|$.
The $\lambda_{\vartheta}<3$ bound was added for improved visibility.}
\end{figure}

Figure~\ref{fig:domains_J1} shows the two-dimensional projections of the
physical parameter domain for inclusive observations ($\lambda_{\vartheta}$,
$\lambda_{\varphi}$, $\lambda_{\vartheta \varphi}$, $A_{\vartheta}$ and
$A_{\varphi}$), in the most general case and in the specific case of
fermion-antifermion decays.

%

\section{\boldmath The polar projection of the $J=2$ case} \label{sec:J2}

The polar projection of the two-body decay distribution of a $J=2$ particle is
described by four independent parameters (defined by
Eq.~\ref{eq:polar_ang_distr_anyJ}),
\begin{align}
\lambda_1 & =   4 (\alpha^-_{22} - 2 \alpha^-_{11} + 4 \alpha^-_{12}) / D \, , \nonumber \\
\lambda_2 & =   6 (\alpha^+_{22} - \alpha^+_{00} - 2 \alpha^+_{02}
-2 \alpha^+_{11} + 4 \alpha^+_{01}) / D \, , \label{eq:ang_distr_Jeq2_parameters} \\
\lambda_3 & =   4 (\alpha^-_{22} + 4 \alpha^-_{11} - 4 \alpha^-_{12}) / D \, , \nonumber \\
\lambda_4 & =   (\alpha^+_{22} + 9 \alpha^+_{00} + 6 \alpha^+_{02} + 16
\alpha^+_{11} - 24 \alpha^+_{01} - 8
\alpha^+_{12}) / D \, , \nonumber \\
\mathrm{with}& \; \; D =   \alpha^+_{22} + \alpha^+_{00} + 6 \alpha^+_{02} + 4
\alpha^+_{11} + 8 \alpha^+_{12} \, , \\
\mathrm{and}& \; \; \alpha^\pm_{i,j} \; 
= \; \sigma_{i,j} + \sigma_{j,i} + \sigma_{-i,-j} + \sigma_{-j,-i} \nonumber \\
&\qquad \; \; \pm ( \sigma_{i,-j} + \sigma_{-j,i} + \sigma_{-i,j} + \sigma_{j,-i} ) \, . 
\label{eq:alphas}
\end{align}
\begin{figure}[htb]
\centering
\resizebox{0.65\linewidth}{!}{%
\includegraphics{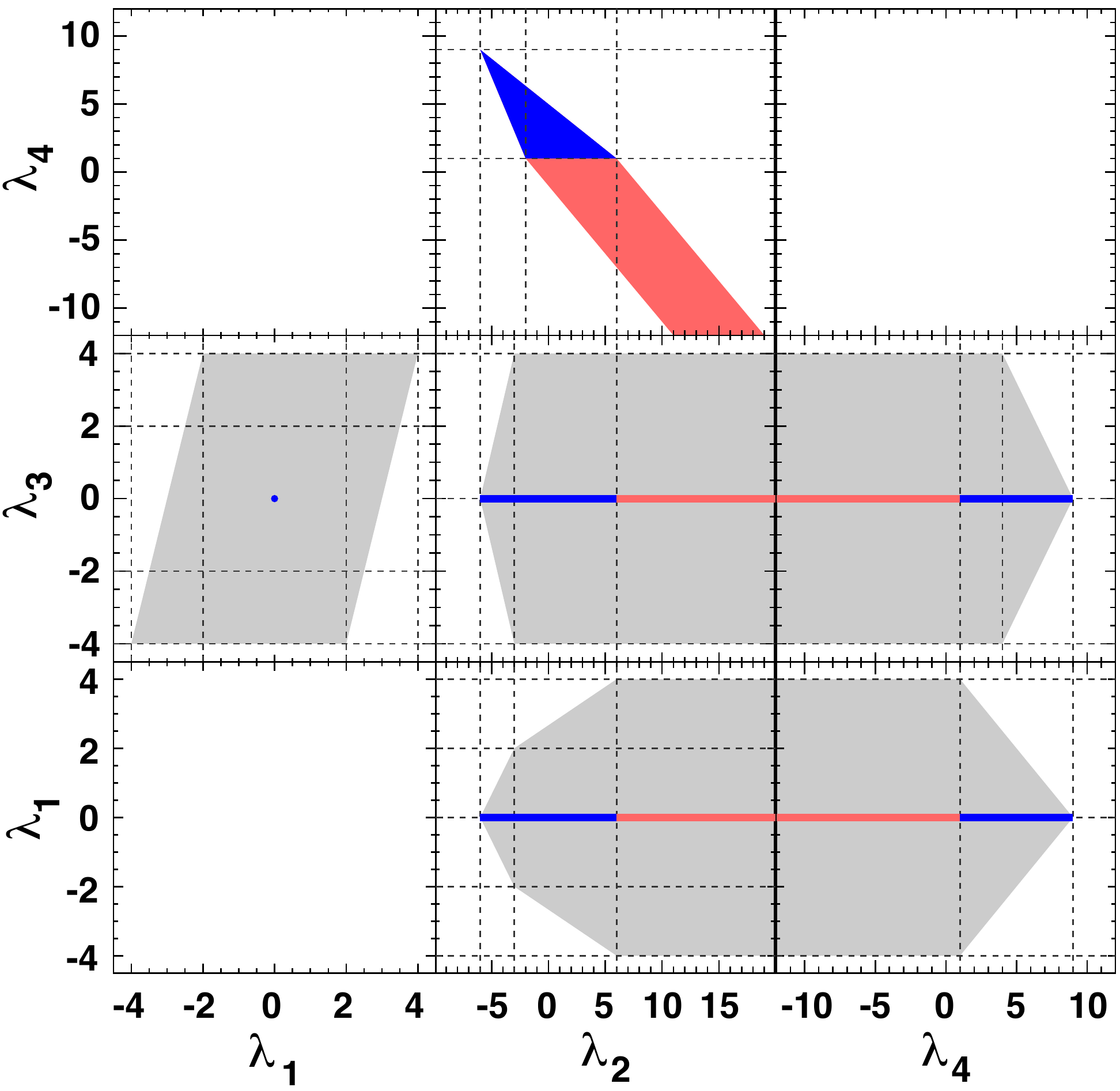}}
\caption{\label{fig:domain_diphoton} Allowed parameter regions of the
polar projection of the two-body decay distribution of a $J=2$ boson. The
largest areas (gray) represent the most general domain. The
intermediate areas (red+blue) represent any of two equivalent cases:
$T$ decays into two real photons; $T$ is produced alone from the scattering of
two real gluons. The smallest areas (blue) represent the case where 
both hypotheses are satisfied. There is no upper bound on $\lambda_2$
and no lower bound on $\lambda_4$.}
\end{figure}
The physical domain of the $\lambda_1, \ldots, \lambda_4$ parameters
depends on the type of decay products and on the production mechanism.
Figure~\ref{fig:domain_diphoton} shows the physically allowed regions of the
$J=2$ parameter space corresponding to MPCs of increasing strength. The largest
areas, in gray, represent the most general case: only angular momentum
conservation and rotation invariance are imposed. The more restricted areas,
in colour, 
represent the effects of selecting one of two specific
physics hypotheses: either the decay products are real photons
($\sigma_{m,\pm1}=0$ for any $m$), or the decaying boson is produced alone from
the scattering of two real (transversely polarized) gluons ($\sigma_{\pm1,m}=0$
for any $m$ when the polarization axis is chosen along the scattering direction
of the gluons). The smallest areas, in blue, show the particularly
interesting case when \emph{both} of these hypotheses apply: the boson is
produced in gluon-gluon fusion and decays into two photons.

\begin{figure}[tb]
\centering
\resizebox{0.65\linewidth}{!}{%
\includegraphics{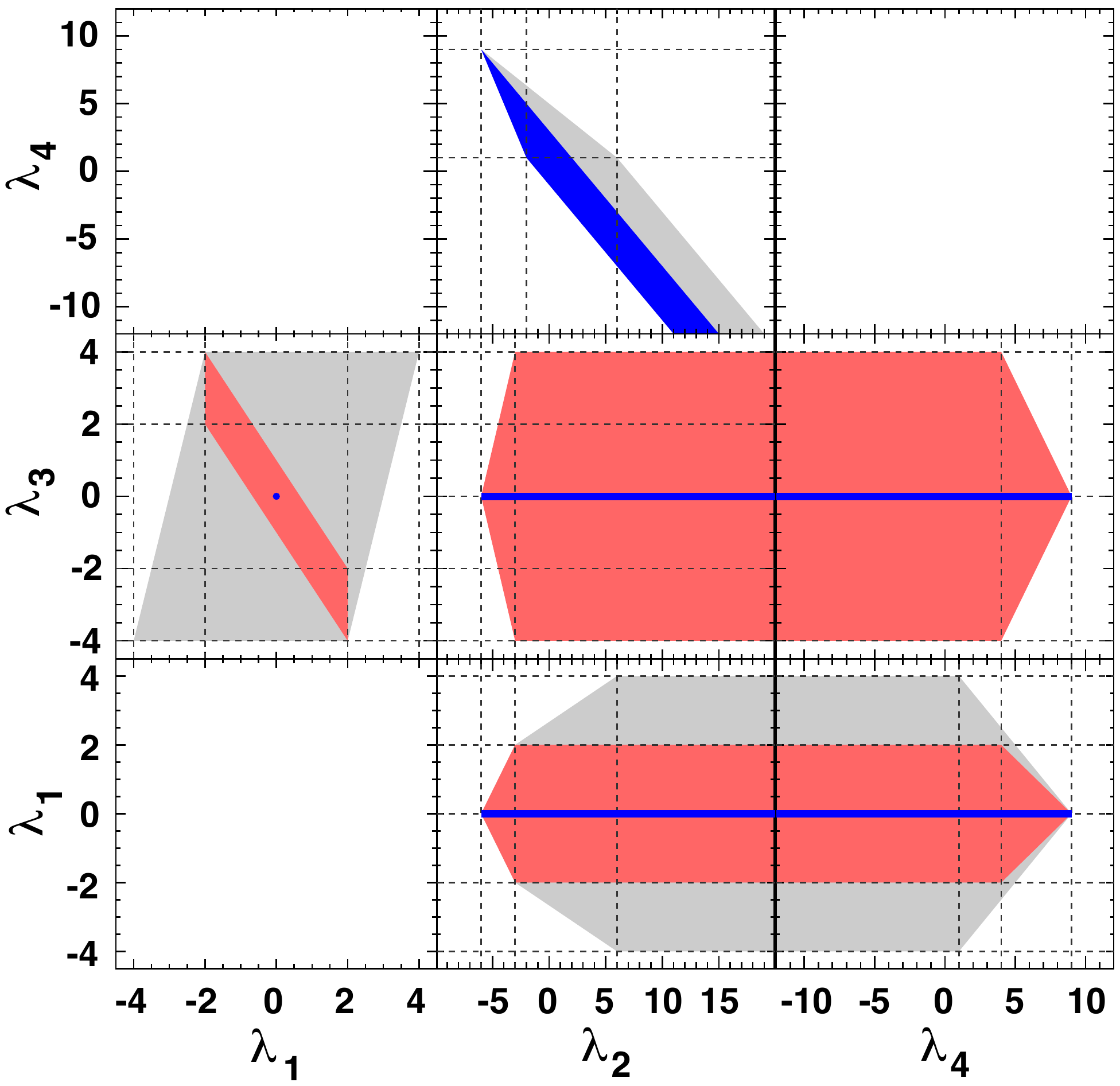}}
\caption{\label{fig:domain_dilepton} Allowed parameter regions of the
polar projection of the two-body decay distribution of a $J=2$ boson. The
largest areas (gray) represent the most general domain. The
intermediate areas (red+blue) represent the case in which $T$ decays into two
$J=1/2$ fermions. The smallest areas (blue) correspond to the hypothesis that
$T$ is produced alone from the scattering of two real gluons \emph{and} decays
into two fermions. There is no upper bound on $\lambda_2$
and no lower bound on $\lambda_4$.}
\end{figure}

Figure~\ref{fig:domain_dilepton} shows two other interesting physical cases.
The intermediate areas, in colour, 
represent the case in which the
boson decays into two $J=1/2$ fermions, while the smallest areas, in
blue, also impose the extra condition that the boson is produced alone from the
scattering of two real gluons.
When either the elementary production process is initiated by two identical
particles or the decay products are two identical particles, the
parity-violating terms $\lambda_1$ and $\lambda_3$ obviously vanish, resulting
in simple lines or dots as allowed regions in the two-dimensional projected
domains.
%

\section{\boldmath Spin characterization of a heavy di-photon resonance} \label{sec:J_eq_0_verification}

Hypotheses on the properties of the fundamental couplings involved can,
obviously, further restrict the allowed regions or even fix the values of the
$\lambda_i$ parameters for the $J=2$ particle. For example, considering the
gluon-gluon to photon-photon case (where $\lambda_1 = \lambda_3 = 0$), the
hypothesis of a graviton-like $J=2$ particle interacting with SM
bosons with no helicity flip~\cite{bib:Gao_etal} (corresponding to the
conditions $\sigma_{0,m}= \sigma_{m,0}=0$ for any $m$) leads to $\lambda_2 = 6$
and $\lambda_4 =1$, the rightmost vertex of the blue triangle in the
$\lambda_2$--$\lambda_4$ plane, top panel of
Fig.~\ref{fig:domain_diphoton}. This particular model has been 
considered as a possible interpretation of the ``Higgs-like
particle" recently discovered by the ATLAS and CMS experiments~\cite{bib:ATLAS,bib:CMS}. 
We will now analyze this physical example as an
illustration of the usefulness of the MPCs.

To determine the angular momentum quantum number of the new
resonance, it is interesting to note that a particle produced (alone) from the
scattering of two real gluons and decaying into two real photons always decays
with a significant polar anisotropy ($\vec{\lambda} \ne 0$) with respect to the
scattering direction of the gluons, as long as $J \ne 0$. This completely
model-independent result, shown in Fig.~\ref{fig:domain_diphoton} for the $J=2$
case ($\lambda_4$ is always $\ge 1$), can be generalized to other $J$ values.
Since the $J=1$ case is excluded by the Landau--Yang theorem, let us consider
$J=3$ and $J=4$.

The polar projection of the di-photon decay distribution of a $J=3$
particle produced by gluon-gluon fusion is described by the
parameters
\begin{align}
\begin{split}
\lambda_2 & =   3 (  3 \alpha^+_{00} + 10 \alpha^+_{02} - 5 \alpha^+_{22} ) / D \, , \\
\lambda_4 & =  10 ( -3 \alpha^+_{00} -  6 \alpha^+_{02} +   \alpha^+_{22} ) / D \, , \\
\lambda_6 & =     ( 25 \alpha^+_{00} + 30 \alpha^+_{02} + 9 \alpha^+_{22} ) / D \, , \\
\mathrm{with}& \; \; D =   4 \alpha^+_{22} \, .
\label{eq:ang_distr_Jeq3_parameters}
\end{split}
\end{align}
The corresponding $J=4$ decay parameters are
\begin{align}
\begin{split}
\lambda_2 & =    4 (-45 \alpha^+_{00} -160 \alpha^+_{02} +52 \alpha^+_{22} ) / D \, , \\
\lambda_4 & =   10 (111 \alpha^+_{00} +312 \alpha^+_{02} -32 \alpha^+_{22} ) / D \, , \\
\lambda_6 & = -140 ( 15 \alpha^+_{00} + 32 \alpha^+_{02} + 4 \alpha^+_{22} ) / D \, , \\
\lambda_8 & =   49 ( 25 \alpha^+_{00} + 40 \alpha^+_{02} +16 \alpha^+_{22} ) / D \, , \\
\mathrm{with}& \; \;
        D =   9 \alpha^+_{00} + 40 \alpha^+_{02} +16 \alpha^+_{22} \, .
\label{eq:ang_distr_Jeq4_parameters}
\end{split}
\end{align}
\begin{figure}[t]
\centering
\resizebox{0.65\linewidth}{!}{%
\includegraphics{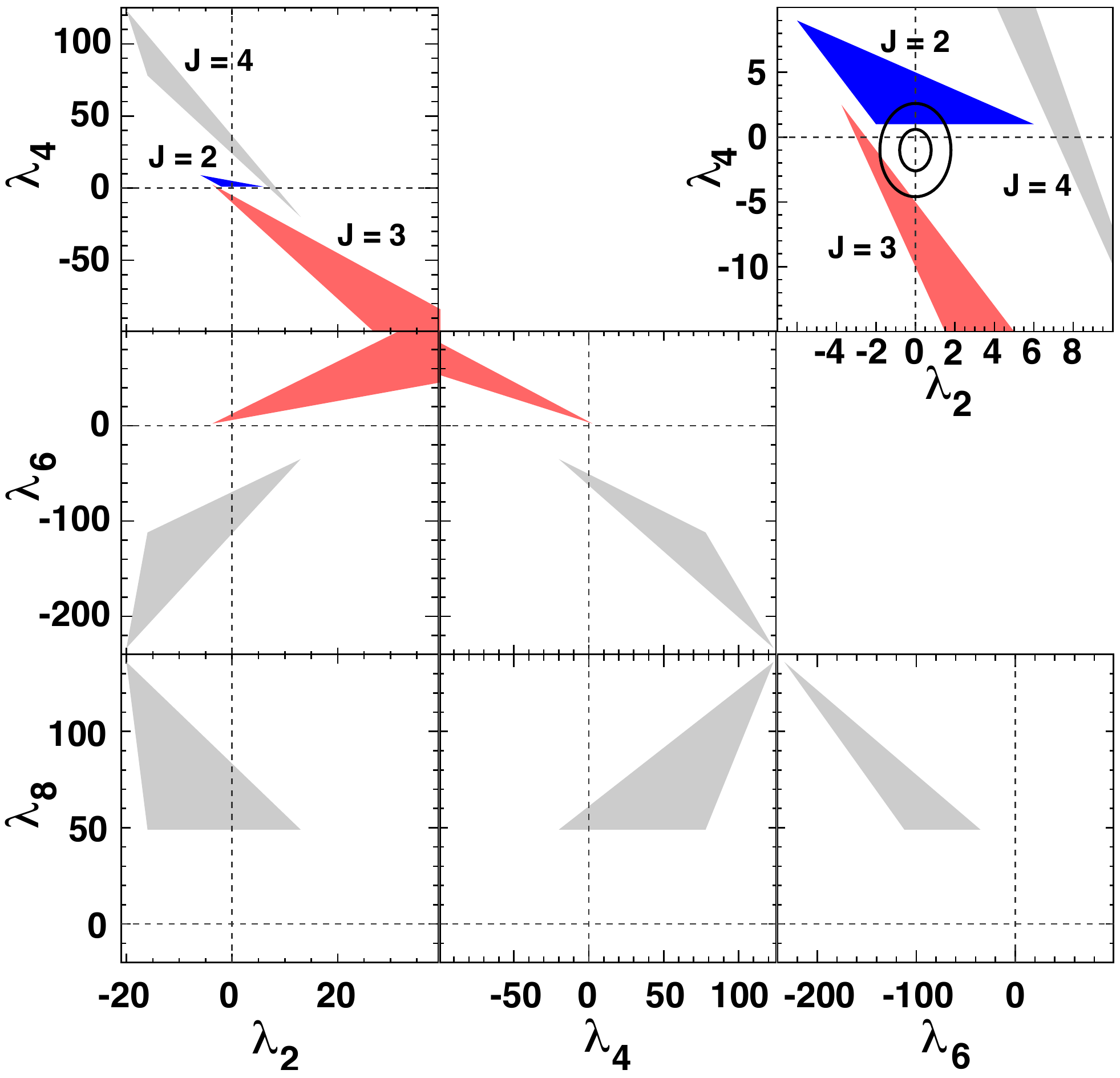}}
\caption{\label{fig:domain_diphoton_J234} Allowed regions for the parameters of
the polar projection of the decay distribution of a particle produced by
gluon-gluon fusion and decaying into two photons. The three different shades
indicate, from darkest (blue) to lightest (gray), the $J=2$, $3$ and $4$ cases.
In the $J=3$ case there are no upper bounds on
$\lambda_2$ and $\lambda_6$ and no lower bound on $\lambda_4$.
The top-right panel shows a zoom of the top-left panel.}
\end{figure}
In both cases the parity-violating $\lambda_i$ parameters (odd $i$) vanish. 
The parameter domains for $J=2$, $3$ and $4$ are shown in
Fig.~\ref{fig:domain_diphoton_J234}. 
The minimum distance from the origin
($\vec{\lambda} = 0$, corresponding to $J=0$) 
increases from $J=2$ to $J=3$ to $J=4$ and,
in general, should
increase
with $J$, reflecting the 
stronger polarization imposed by the 
limitation of the initial- and final-state helicities to $|m| \le 2$ and
$|k^\prime| \le 2$.
%
%

It is interesting to note that the three domains have no intersections between
them and also not with the $J=0$ point ($\vec{\lambda} = 0$). Therefore, a
sufficiently-precise measurement of the di-photon decay distribution can
provide an unambiguous spin characterization, independent of specific
hypotheses on the identity of the particle.

The experimental precision needed to achieve a significant discrimination
obviously depends on $J$ and on the actual values of the polarization
parameters, defined by the identity of the particle.
The top-right panel of
Fig.~\ref{fig:domain_diphoton_J234} includes two
ellipses that illustrate two putative measurements, the radii representing 
their uncertainties.
The outer one excludes the $\lambda_2 = 6$, 
$\lambda_4 =1$ point, eliminating the hypothesis that the decaying boson is
a graviton-like $J=2$ particle of the kind discussed in Ref.~\cite{bib:Gao_etal}.
However, it does not rule out other hypothetical $J=2$ (or even $J=3$) bosons,
corresponding to parameters (calculable from the relevant couplings) 
closer to the $\lambda_2 = \lambda_4 = 0$ origin (that represents the 
isotropic distribution of a $J=0$ particle).
The inner ellipsis illustrates another hypothetical measurement, sufficiently
precise to exclude the full $J=2$ and $J=3$ domains, thereby leading 
to the model-independent spin characterization of a $J=0$ particle.

\begin{figure}[tb]
\centering
\includegraphics[width=0.48\linewidth]{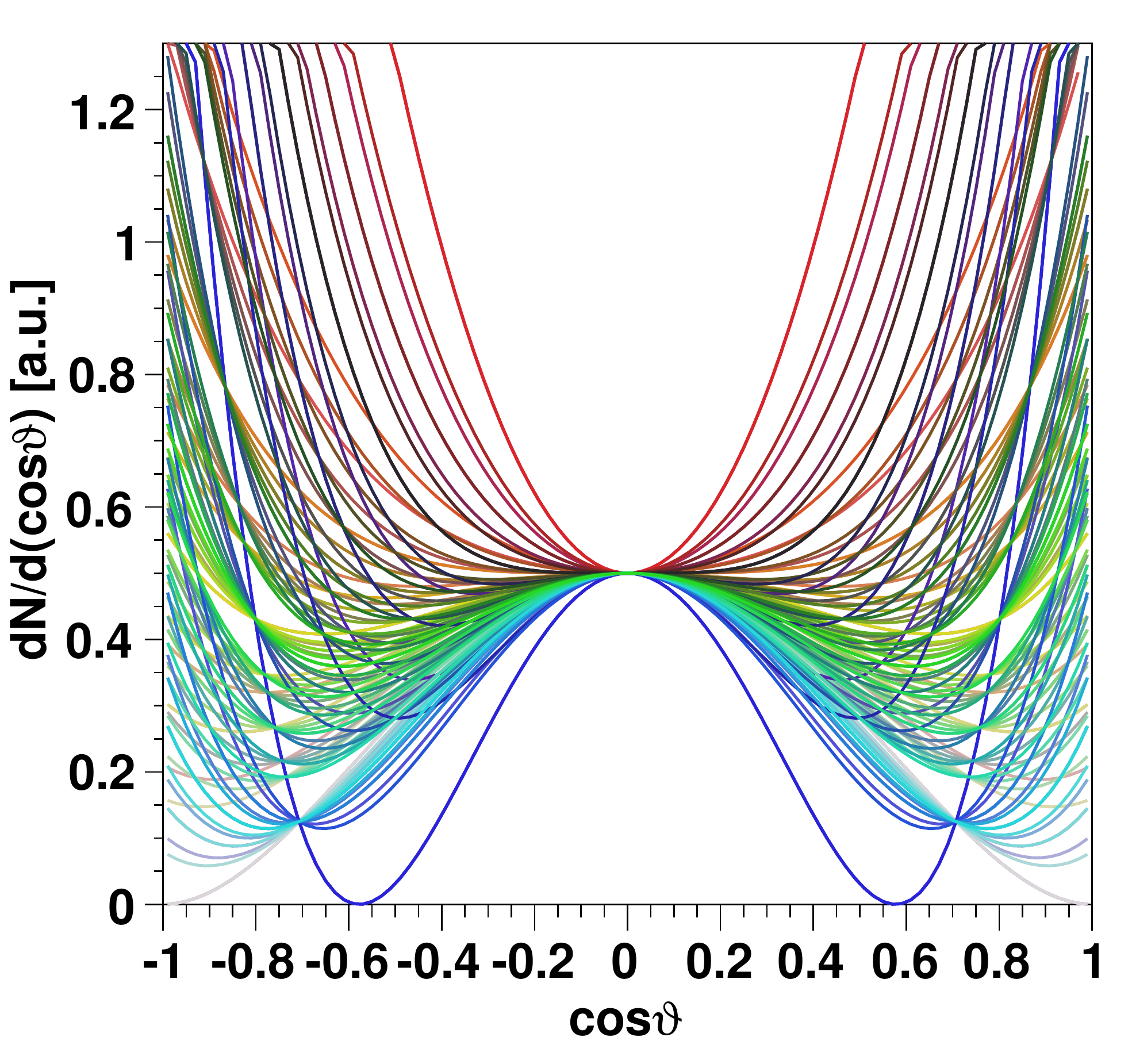}
\includegraphics[width=0.48\linewidth]{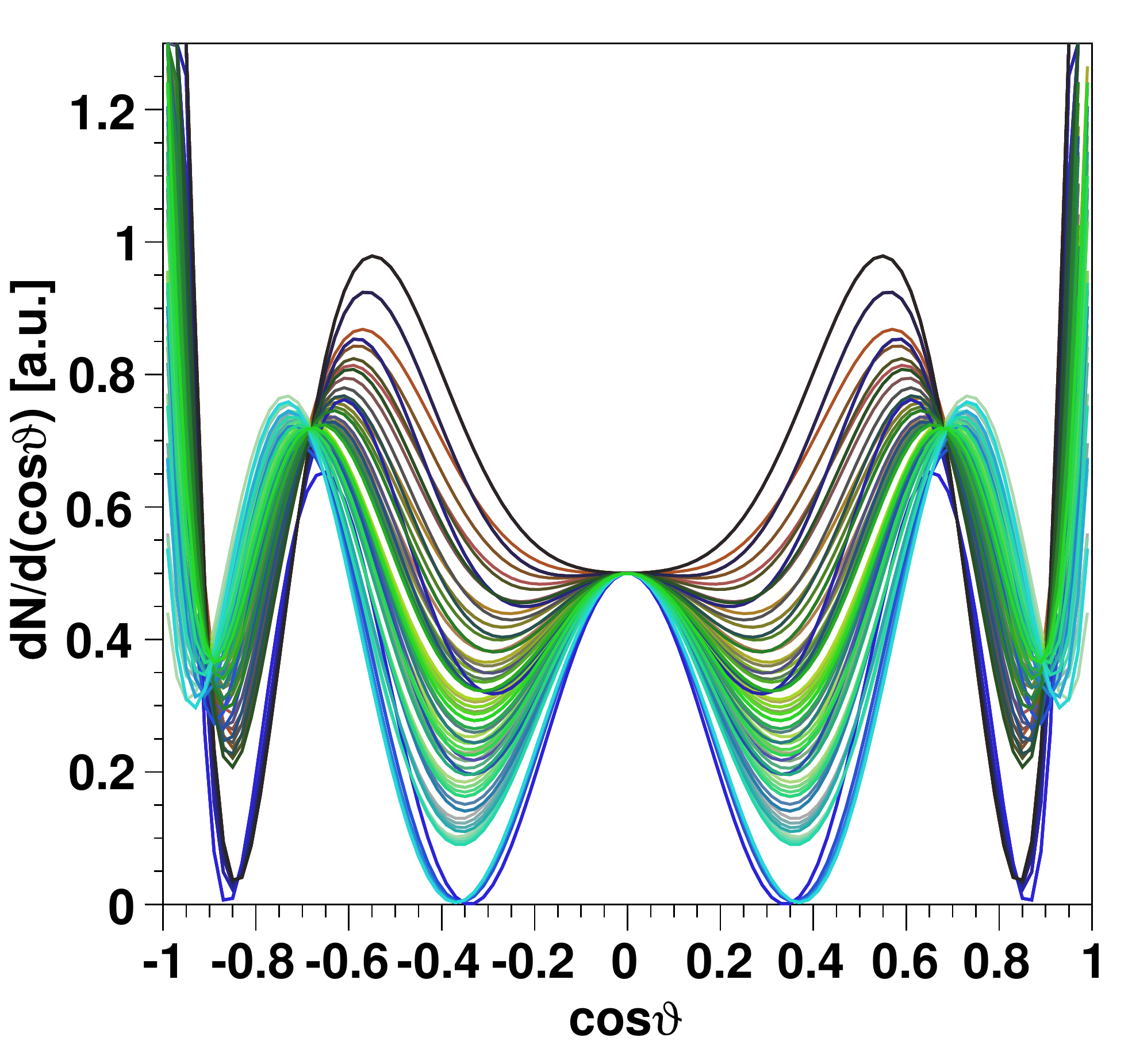}
\caption{\label{fig:J234_diphoton_costheta} Physically allowed
$\cos\!\vartheta$ distributions for $J=2$ (left) and $J=4$ (right) bosons
produced by gluon-gluon fusion and decaying into two photons. For
clarity of representation, all distributions are normalized to the same value
at $\cos\!\vartheta = 0$.}
\end{figure}

Figure~\ref{fig:J234_diphoton_costheta} illustrates the dependence of the
observable $\cos\!\vartheta$ distribution on $J$, taking as examples the $J=2$
and $4$ cases. The curves were obtained by scanning the (respectively, two- and
four-dimensional) physical domains of the $\lambda_i$ parameters. The 
recognizable difference in shape between any single $J=2$ curve and any single
$J=4$ curve shows that it is always possible, with a
sufficiently-precise measurement, to unambiguously determine one and only one
spin value.

As shown in Fig.~\ref{fig:domain_diphoton_J234}, the physically-allowed 
parameter regions of the $J=0, 2, 3$ and $4$ cases do not intersect each
other, a non-trivial observation that can potentially lead to
significant improvements in some of the LHC data analyses. 
However, it should be kept in mind that these parameter regions
have been established under the (commonly-assumed) condition that the boson
under study is produced in a $2 \to 1$ (leading order) process. While one can
often argue (in the context of differential cross sections, for instance) that
next-to-leading order contributions can be neglected in first approximation,
this might not be defendable in studies aimed at measuring the spin of a new
resonance. Indeed, from a polarization perspective, the $2 \to 1$ and $2 \to 2$
cases are completely different. Even if the extra emitted particles have almost
zero momentum, the simple symmetry of the $2 \to 1$ reaction is broken and
the polarization, which is a topological property, changes in a drastic way. In
the $J=2$ case, for instance, there would no longer be any restriction of the
most general parameter domain, as can be seen in Fig. 3 (top panel), where
the most general $\lambda_2$ vs.\ $\lambda_4$ region (gray) coincides with the
``intermediate region'' (red plus blue). The most constrained parameter space
(blue), which, in particular, excludes the unpolarized option, is only obtained
when we complement the two-photon decay restriction with the $2 \to 1$
production hypothesis.

\section{Concluding remarks} \label{sec:summary}

A complete knowledge of how the parameters of a decay angular distribution are
constrained and mutually correlated simply by rotation invariance, angular
momentum conservation and the identities of the initial- and final-state
particles (i.e., with minimal or no hypotheses on the nature, properties and/or
polarization of the decaying particle itself) can bring important benefits to
experimental analyses. In polarization measurements, significant violations
of such MPCs (in this case defined as constraints not imposing any condition on
the polarization itself) can indicate the existence of systematic biases in the
analysis, related, for example, to the acceptance determination or to the
modelling and subtraction of the backgrounds.

Moreover, the analysts can (and should) embed the prior knowledge of the
physical bounds into the measurement, according to the Bayesian reasoning. This
procedure can appreciably change the significance of the
result when the observation is interestingly at the border of the parameter
space. This is usually the case, for example, of the SM vector
gauge bosons, strongly polarized (either transversely or longitudinally) both
when produced directly in parton-parton scattering (Drell--Yan-like processes)
and when coming from the decay of massive particles (top quark, heavy Higgs
boson).

Finally, the physical parameter domain for a given process depends on the spin
of the decaying particle. Its detailed knowledge can, therefore, provide a
model-independent instrument for the spin characterization of newly discovered
particles. This approach may require larger acquired event samples with respect
to binary comparisons between specific physical hypotheses, but provides a
more model-independent and unequivocal answer to the basic physical question.

\bigskip

P.F.\ and J.S.\ acknowledge support from Funda\c{c}\~ao para a
Ci\^encia e a Tecnologia, Portugal, under contracts
SFRH/BPD/42343/2007 and CERN/FP/109343/2009.


\end{document}